\documentclass[prb,twocolumn,groupedaddress,amsfont,showpacs]{revtex4}
\usepackage{amssymb}

\usepackage{graphicx}
\usepackage{amsmath}
\usepackage{amssymb}

\begin{document}

\title{Effect of temperature on resonant electron transport through stochastic conduction channels in superlattices.}

\author{A.O. Selskii$^1$, A.A. Koronovskii$^1$, A.E.
Hramov$^1$, O.I. Moskalenko$^1$, K.N. Alekseev$^2$, M.T. Greenaway$^3$,  T.M.
Fromhold$^3$, A.V. Shorokhov$^4$, N.N. Khvastunov$^4$ and  A.G. Balanov$^{1,2}$}
\affiliation{$^1$Faculty of Nonlinear Processes, Saratov State University,
Astrakhanskaya, 83, Saratov, 410012, Russia\\
$^2$Department of Physics, Loughborough University, Loughborough LE11 3TU, United Kingdom \\
$^3$School of Physics and Astronomy, University of Nottingham, Nottingham NG7 2RD, United Kingdom\\
$^4$ Institute of Physics and Chemistry, Mordovian State University, 430005 Saransk, Russia}

\date{\today}

\begin{abstract}
We show that resonant electron transport in semiconductor superlattices with an applied electric and tilted magnetic field can, surprisingly, become more pronounced as the lattice and conduction electron temperature increases from 4.2 K to room temperature and beyond. It has previously been demonstrated that at certain critical field parameters, the semiclassical trajectories of electrons in the lowest miniband of the superlattice change abruptly from fully localised to completely unbounded. The unbounded electron orbits propagate through intricate web patterns, known as stochastic webs, in phase space, which act as conduction channels for the electrons and produce a series of resonant peaks in the electron drift velocity versus electric field curves. Here, we show that increasing the lattice temperature strengthens these resonant peaks due to a subtle interplay between thermal population of the conduction channels and transport along them. This enhances both the electron drift velocity and the influence of the stochastic webs on the current-voltage characteristics, which we calculate by making self-consistent solutions of the coupled electron transport and Poisson equations throughout the superlattice. These solutions reveal that increasing the temperature also transforms the collective electron dynamics by changing both the threshold voltage required for the onset of self-sustained current oscillations, produced by propagating charge domains, and the oscillation frequency.
\end{abstract}



\maketitle

\section{Introduction}

Semiconductor superlattices (SLs) are nanostructures formed from several alternating layers of different semiconductor materials \cite{Esaki:1970_SuperLattices,SHI75,WAC02, Tsu:2005_Superlattice_book}. This periodic structure leads to the formation of energy minibands that enable electrons, in the presence of an electric field, to demonstrate a number of interesting quantum-mechanical phenomena, which include the formation of Wannier-Stark ladders, sequential and resonant tunneling, Bragg reflections, and Bloch oscillations. Consequently, SLs are of a great interest for both fundamental and applied science \cite{Esaki:1970_SuperLattices,Romanov72,IGN87,mendez,Holthaus,ZHA96,SHOM98,AMA02,HIZ06,HYART07,END07,HYART09}. Due to the high mobility of miniband electrons and the very high frequency of the Bloch and charge-domain oscillations, SLs have prospective applications in sub-THz and THz electronic devices  \cite{ZHA96,SHOM98,END07,KHOS09,Greenaway10}.

Recently, it has been shown that a tilted magnetic field applied to a SL can strongly affect, and hence control, the electrical properties of the structure. Nonlinear interaction between the electronic Bloch oscillations along the superlattice and cyclotron motion in the plane of the layers induces chaotic semiclassical electron dynamics, which, depending on the ratio between the Bloch and cyclotron frequencies, either accelerate or decelerate charge transport through the SL \cite{From01,From04}. Coupling of cyclotron and Bloch motion by a tilted magnetic field has also been related to the Fiske effect in superconducting Josephson junctions \cite{KOS06}. On resonance, when the Bloch and cyclotron frequencies are  commensurate, the electrons exhibit a unique type of quantum chaos, which does not obey Kolmogorov-Arnold-Moser theory\cite{Sag88}. This type of chaos is characterised  by the formation of intricate ``web-like'' structures, known in the literature as ``stochastic webs'' \cite{Sag88,Zasl91,Buch}, which extend throughout the phase space of the miniband electrons. The appearance of these webs abruptly delocalises the electrons in real space, thus significantly increasing their drift velocity \cite{From01,Balanov:2008_SL-PRE}. Similar dynamics can occur in other spatially periodic systems in which wave interference gives rise to band transport phenomena, including ultracold atoms in an optical lattice \cite{SCO02}, graphene \cite{Ferry}, and light propagating through spatially-modulated photonic crystals \cite{WIL03,Henn11}. On the quantum level, such resonant delocalisation of the electrons manifests itself in the formation of additional magnetic-field-induced miniband structure corresponding to extended electron states, which extend across many periods of the SL \cite{Fowler07}. This delocalisation dramatically affects the collective electron behavior by inducing multiple propagating charge domains and GHz-THz current oscillations with frequencies and amplitudes much higher than with no tilted field \cite{Greenaway:2009_SL_B}. It has also been shown that in the vicinity of Bloch-cyclotron resonances the usually unstable Bloch gain profile becomes stable \cite{Hyart09}, which can be used for the amplification of THz signals. Although the above effects of the Bloch-cyclotron resonances on electron dynamics have been reported for different finite temperatures, it is still unclear how the character of charge transport in the SL changes with temperature.
\begin{figure}[t]
  \centering
\includegraphics*[width=.8\linewidth]{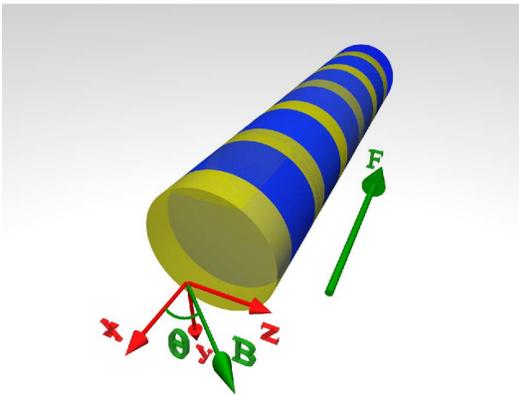}
  \caption{(Color online) Schematic diagram of a semiconductor superlattice with an electric field and a tilted magnetic field applied. Red (green) arrows indicate co-ordinate axes (field orientation). Blue (yellow) bands in the $y-z$ plane represent quantum wells (tunnel barriers) within the superlattice. Electric field, $\mathbf{F}$, is applied anti-parallel to the $x$-axis and the magnetic field, $\mathbf{B}$, lies in the $(x,z)$-plane at an angle $\theta$ to the $x$-axis.} 
\label{fig:conf}
\end{figure}

In this paper, we study how increasing the electron and lattice temperature affects the drift velocity of the electrons in an electric and tilted magnetic field and, consequently, the electric current through the SL. We find that the thermal distribution of the electrons can, counter-intuitively, enhance the effect of Bloch-cyclotron resonances on the drift velocity. In particular, increasing the temperature makes the resonant maxima in the drift velocity versus electric field characteristics more prominent. This shifts the frequency and amplitude of the collective charge-domain oscillations, associated with each resonant peak, together with the threshold voltage at which the oscillations switch on. 

\section{Electron drift velocity}
\label{sec:Sec1}

We consider the field configuration shown schematically in Fig. \ref{fig:conf}. The electric field  $\textbf{F}=(-F, 0, 0)$ is applied perpendicular to the plane of the SL layers and anti-parallel to the $x$-axis. The magnetic field $\textbf{B}=(B \cos \theta, 0, B \sin \theta)$ lies in the $(x,z)$-plane at an angle $\theta$ to the $x$-axis. The semiclassical equation of motion for a miniband electron is
\begin{equation}
  \dot{\mathbf{p}} (t) = - e [ \mathbf{F} + (
  \mathcal{\nabla_{\mathbf{p}}} E ( \mathbf{p} (t)) \times
  \mathbf{B})] \label{veceq}
\end{equation}
where $e$ is the magnitude of the electron charge, $\mathbf{p}(t)=(p_x(t),p_y(t),p_z(t))$ is the electron's crystal momentum at time $t$ and $E(\mathbf{p})=\Delta (1-\cos(p_x d/\hbar))/2 +(p^2_y+p^2_z)/2m^{\ast}$ is the dispersion relation for the lowest miniband within the tight-binding approximation. Here, $\Delta$ is the miniband width, $d$ is the SL period, and $m^{\ast}$ is the electronic effective mass for motion in the $(y,z)$-plane.

Equation (\ref{veceq}) can be expressed in its constituent components  \cite{From01,Balanov:2008_SL-PRE}
\begin{eqnarray}
  \dot{p}_x (t) & = & eF - \omega_{\perp} p_y (t)
  \label{compeq1}\\
  \dot{p}_y (t) & = & \frac{d \Delta m^{\ast}
  \text{$\text{$\omega_{\perp}$}$}}{2 \hbar} \sin \left( \frac{p_x (t)
  d}{\hbar} \right) - \text{$\text{$\text{$\omega_{\|}$}$}$} p_z (t)
  \label{compeq2}\\
  \dot{p}_z (t) & = & \text{$\text{$\omega_{\|}$}$} p_y (t),  \label{compeq3}
\end{eqnarray}
where $\omega_{\|}$=$eB \cos \theta / m^{\ast}$ and $\omega_{\perp}$=$eB
\sin \theta / m^{\ast}$ are the cyclotron frequencies corresponding to the
magnetic field components along the $x$- and $z$-axes, respectively. The electron velocity along the $x$-direction is given by
\begin{eqnarray}
  v_x(t)=\dot{x}(t)= v_0 \sin \left( \frac{p_x (t) d}{\hbar} \right), \label{eq:vx}
\end{eqnarray}
where the peak velocity $v_0=\Delta d/(2\hbar)$.

To determine the drift velocity, $u_d$, of electrons with initial momentum $\mathbf{P}=(P_x,P_y,P_z)$ we use the Esaki-Tsu formalism \cite{Esaki:1970_SuperLattices,From04}:
\begin{eqnarray}
u_d(\mathbf{P})=\nu \int_0^{\infty} v_x(t) e^{-\nu t} dt, \label{eq:ud}
\end{eqnarray}
where $\nu$ is the electron scattering rate \cite{footnote1}. 

In the case of non-zero temperature, $T$, one should take into account the thermal distribution of electron momenta $f(P_x,P_y,P_z)$, which we assume to obey the Boltzmann statistics \cite{Romanov72,Bass80,Bass86}
\begin{eqnarray}
f (\mathbf{P}) = \frac{1}{Z} e^{- \frac{\Delta}{2 k_B T}
   \left( 1 - \cos \frac{P_x d}{\hbar} \right)
   - \frac{P^2_y + P^2_z}{2m^{\ast} k_B T} }. \label{eq:Boltzmann}
\end{eqnarray}
In Eq. (\ref{eq:Boltzmann}), $Z$ can be found from the normalisation condition
 \begin{eqnarray}
\int_{-\pi \hbar/d}^{\pi \hbar/d}\int_{-\infty}^{\infty}\int_{-\infty}^{\infty}
f(\mathbf{P})d P_x dP_y dP_z =1, \nonumber
\end{eqnarray}
which yields
\begin{equation}
  Z = (2 \pi)^2 m^{\ast} k_B T \frac{\hbar}{d} I_0 \left( \frac{\Delta}{2 k_B
  T} \right) e^{- \frac{\Delta}{2 k_B T}}, \label{eq:Znorm}
\end{equation}
where $I_0 (x)$ is a modified Bessel function of the first kind.
Then, the drift velocity of miniband electrons at temperature $T$ is
\begin{eqnarray}
v_d = \int_{-\pi \hbar/d}^{\pi \hbar/d}\int_{-\infty}^{\infty}\int_{-\infty}^{\infty} f(\mathbf{P})u_d(\mathbf{P}) d P_x dP_y dP_z. \label{eq:vdT}
\end{eqnarray}

In the Appendix we show, following previous work \cite{Bass80,Bass86}, that for small angles $\theta$, for which $\omega_{\perp}\ll \omega_{\|}$ and $\omega_{\perp}\ll \omega_B$, where $\omega_B=e F d/\hbar$ is the frequency of the Bloch oscillations, the drift velocity can be approximated by
\begin{widetext}
\begin{eqnarray}
v_d = v_0 \frac{I_1 (\Delta / 2 k_B T)}{I_0 (\Delta / 2 k_B T)} \exp \left[
   - m^{\ast} k_B T \left( \frac{\omega_{\perp}}{\text{$\omega_{\|}$}}
   \frac{d}{\hbar} \right)^2 \right]
  \sum^{\infty}_{n = - \infty} I_n \left[
   m^{\ast} k_B T \left( \frac{\omega_{\perp}}{\text{$\omega_{\|}$}}
   \frac{d}{\hbar} \right)^2 \right] \frac{\nu (\omega_B - n
   \omega_{\|})}{\nu^2 + (\omega_B - n \omega_{\|})^2}, \label{eq:Bass}
\end{eqnarray}
\end{widetext}
where $I_n(x)$, $n=1,2\ldots$ are the modified Bessel functions of the first kind.
When $B=0$, Eq. (\ref{eq:Bass}) reduces to
 \begin{eqnarray}
v_d = v_0 \frac{I_1 (\Delta / 2 k_B T)}{I_0 (\Delta / 2 k_B T)}
  \frac{\nu \omega_B}{\nu^2 + \omega_B^2}. \label{eq:Romanov}
\end{eqnarray}

Note that Eq. (\ref{eq:Romanov}) can be derived exactly \cite{Romanov72} by integrating Eqs. (\ref{compeq1})-(\ref{compeq3}) with $B=0$, and substituting the resulting expressions for $p_x$, $p_y$ and $p_z$ into Eqs. (\ref{eq:vx}) and (\ref{eq:vdT}). For $T=0$ the equation (\ref{eq:Romanov}) yields the famous Esaki-Tsu relation for the electron drift velocity \cite{Esaki:1970_SuperLattices}.

In the general case of arbitrary parameters $F$, $B$ and $\theta$, Eqs. (\ref{compeq1})-(\ref{compeq3}), which, in principle, can exhibit deterministic chaos \cite{From01,From04,Balanov:2008_SL-PRE}, cannot be solved analytically, and therefore the electron drift velocity must be calculated numerically.
In our computations we use the following SL parameters, taken from recent experiments \cite{Pat02,From04}: $d=8.3$ nm, $\Delta=19.1$ meV, $\nu=4\times10^{12}$ s$^{-1}$ and $m^{\ast}=0.067m_e$, where $m_e$ is the mass of a free electron.

\section{Numerical calculation of the electron drift velocity}
\label{sec:num_proc}

In our numerical simulations, we integrate Eqs. (\ref{compeq1})-(\ref{compeq3}) for a number of different initial conditions $\mathbf{p}(0)=\mathbf{P}$, which were randomly generated according to the probability density function (\ref{eq:Boltzmann}). This function can be decomposed into momentum components
\begin{eqnarray}\label{eq:distr_decomp}
f(\mathbf{P})&=&\frac{1}{\sqrt{2\pi m^*k_B T}}
\displaystyle \exp \left(-\displaystyle \frac{\displaystyle P_y^2}{\displaystyle 2m^*k_B T} \right)\\ \nonumber
&\times&\frac{1}{\sqrt{2\pi m^*k_B T}}\displaystyle \exp \left(-\displaystyle \frac{\displaystyle P_z^2}{\displaystyle 2m^*k_B T} \right) \\
&\times&\frac{\displaystyle d}{\displaystyle 2\pi\hbar I_0\left(\frac{\Delta}{2kT}\right)}
\displaystyle \exp \left(\frac {\Delta}{2k_B T}\cos\left(\frac{P_x d}{\hbar}\right)
\right). \nonumber
\end{eqnarray}

As a consequence of this decomposition, one can use independent random number generators for each of the initial momentum components $P_x$, $P_y$, $P_z$. The values of $P_y$ and $P_z$ can be obtained from standard routines \cite{Press_160} for the generation of uncorrelated random numbers obeying a Gaussian distribution with zero mean and variance $(k_BTm^{\ast})^{1/2}$. However, the generation of random values of $P_x$ requires a more sophisticated procedure, which we now explain.

Let us find a continuous function, $g(\xi)$, which transforms a random variable, $\xi$, uniformly distributed within the interval $[-\pi,\pi)$,  into a random variable, $g$, having a distribution
\begin{equation}\label{eq:pxdistr}
f(g) = \frac{\displaystyle d\exp\left(\frac{\Delta}{2k_BT}\right)}{\displaystyle 2\pi\hbar I_0\left(\frac{\Delta}{2k_BT}\right)}
\displaystyle \exp \left(-\frac {\Delta}{2k_BT}\left[1-\cos\left(\frac{g d}{\hbar}\right)\right]
\right).
\end{equation}

To find $g(\xi)$, we use the probability conservation relation
$(1/2\pi)d\xi=f(g)dg$, which gives us a differential equation for $g(\xi)$:
\begin{equation}
\frac{d g(\xi)}{d \xi}=\frac{1}{2\pi f[g(\xi)]}. \label{eq:eq_dis}
\end{equation}

Due to the symmetry of the distribution in Eq. (\ref{eq:pxdistr}), $g(0)=0$, which can be used as an initial condition for Eq. (\ref{eq:eq_dis}). Moreover, it is possible to show that $g(\pm \pi)=\pm \pi$ for any values of temperature $T$. However, in the general case of arbitrary argument, $\xi$, $g(\xi)$ cannot be expressed analytically. We therefore evaluate it by numerical integration of Eq. (\ref{eq:eq_dis}).
\begin{figure}[b]
  \centering
\includegraphics*[width=.75\linewidth]{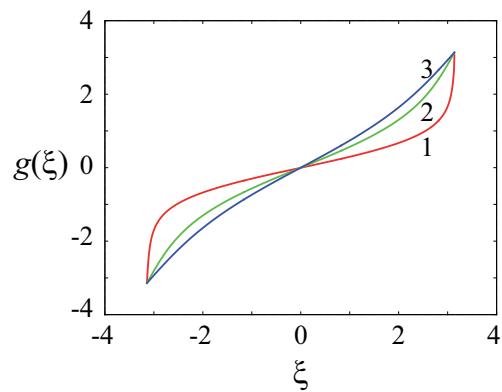}
  \caption{(Color online) Function $g(\xi)$ for three different temperatures: $T=50$ K (curve 1), $T=150$ K (curve 2), $T=300$ K (curve 3). $d=8.3$ nm, $\Delta=19.1$ meV, $\nu=4\times10^{12}$ s$^{-1}$.
\label{fig:transf}}
\end{figure}
Figure \ref{fig:transf} illustrates the function $g(\xi)$ for three different temperatures $T=50$ K (curve 1), $T=150$ K (curve 2), $T=300$ K (curve 3). As the figure shows, $g(\xi)$ is an odd function, which becomes almost linear as $T$ increases.

In practice, numerical integration of Eq. (\ref{eq:eq_dis}) significantly slows the simulations. Therefore, we used approximate analytical solutions. We found that for small temperatures, $T\le50$ K, the distribution (\ref{eq:pxdistr}) can be well approximated by a Gaussian with zero mean and variance $2k_B T \hbar/(\Delta d)$. Hence, in this case $P_x$ can be also defined using the Gaussian random number generator.

For larger temperatures $T>50$ K we approximate the solution of  Eq. (\ref{eq:eq_dis}) by the following polynomial expression
\begin{equation}\label{eq:Poly}
f(\xi) = a_1 \xi + a_3 \xi^3 + a_5 \xi^5 + a_7 \xi^7+O(\xi^9),
\end{equation}
where the coefficients $a_n$ for different temperatures are given in Tab. \ref{tab_1}.
\begin{table}[t]
\caption{Coefficients $a_i$ for the polynomial expansion in Eq. (\ref{eq:Poly}) at different temperatures $T$\label{tab_1}} {
\begin{center}
\begin{tabular}{|c|c|c|c|c|c|}
\hline
~~~$T$ ~~&~$a_1$~~~&~~~$a_3$~~~&~~~$a_5$~~~&~~~$a_7$~~~\\
\hline \hline
~$100 K$~~&~$0.42008$~&~$0.04124$~&~$-0.00755$~&~$0.00094$~\\
~$125 K$~~&~$0.49513$~&~$0.02223$~&~$0.00087$~&~$0.00039$~\\
~$150 K$~~&~$0.55066$~&~$0.01556$~&~$0.00210$~&~$0.00010$~\\
~$175 K$~~&~$0.59097$~&~$0.01440$~&~$0.00309$~&~$-0.00003$~\\
~$200 K$~~&~$0.62431$~&~$0.01505$~&~$0.00329$~&~$-0.00009$~\\
~$225 K$~~&~$0.65325$~&~$0.01626$~&~$0.00312$~&~$-0.00012$~\\
~$250 K$~~&~$0.67807$~&~$0.01744$~&~$0.00282$~&~$-0.00013$~\\
~$275 K$~~&~$0.69868$~&~$0.01836$~&~$0.00249$~&~$-0.00013$~\\
~$300 K$~~&~$0.71677$~&~$0.01905$~&~$0.00217$~&~$-0.00012$~\\
~$325 K$~~&~$0.73367$~&~$0.01954$~&~$0.00186$~&~$-0.00011$~\\
~$350 K$~~&~$0.74852$~&~$0.01983$~&~$0.00159$~&~$-0.00010$~\\
~$375 K$~~&~$0.76150$~&~$0.01996$~&~$0.00136$~&~$-0.00009$~\\
~$400 K$~~&$0.77299$~&~$0.01998$~&~$0.00116$~&~$-0.00009$~\\
\hline
\end{tabular}
\end{center}
}
\end{table}
Note that with increasing $T$, the coefficient $a_1$, corresponding to the linear term, grows, whereas the other (nonlinear) coefficients, $a_3$, $a_5$, $a_7$,  decrease. Also, the higher-order terms can be neglected.

Thus, to calculate the drift velocity, $v_d$, for given temperature, $T$, we (i) randomly generate the components of the initial momentum vector $\mathbf{P}$, (ii) use them as an initial condition to integrate Eqs. (\ref{compeq1})-(\ref{compeq3}), (iii) substitute the results of this integration into Eqs. (\ref{eq:vx}) and (\ref{eq:ud}) to determine the drift velocity, $v_d$, for the given initial momentum, and (iv) average $v_d$ over all randomly-chosen initial momenta.

To ensure that the above procedure for simulating the thermal distribution of electron momenta is accurate enough, we check it for the case of no applied magnetic field, for which $v_d$ can be calculated analytically \cite{Romanov72} using Eq. (\ref{eq:Romanov}).
\begin{figure}[t]
  \centering
\includegraphics*[width=1\linewidth]{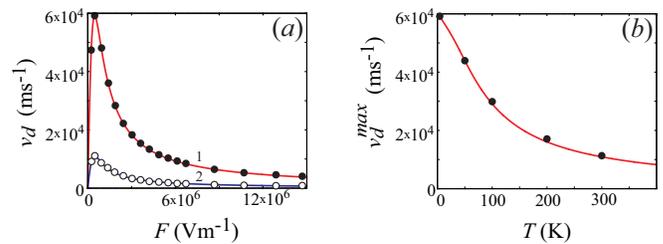}
 \caption{(Color online) (a) Variation of drift velocity, $v_d$, with electric field, $F$, determined analytically from Eq. (\ref{eq:Romanov}) (solid curves) and calculated numerically (symbols) for $T=4.2$ K (curve 1 and ``$\bullet$'' ) and $T=300$ K  (curve 2 and ``$\circ$''). (b) Variation of maximal drift velocity, $v_d^{max}$, with $T$ calculated analytically (solid line) and numerically  ($\bullet$). Numerical simulations involved averaging over 250000 different initial momenta, $\mathbf{P}$, chosen randomly.
\label{fig:vdB0}}
\end{figure}
The solid curves in Fig. \ref{fig:vdB0}(a) shows $v_d(F)$ plots calculated using Eq. (\ref{eq:Romanov}) for $T=4.2$ K (upper curve) and $T=300$ K (lower curve). The data points marked by symbols ``$\bullet$'' and ``$\circ$'' show values of $v_d$ obtained numerically for $T=4.2$ K and $T=300$ K, respectively. The figure demonstrates excellent agreement between the exact Eq. (\ref{eq:Romanov}) and our numerical procedure. More precisely, we find that by averaging over $N \ge 200000$ trajectories, the relative error does not exceed $1 - 2 \%$. Consequently, in all our simulations we used $N=250000$.

 Eq. (\ref{eq:Romanov}) shows that the maximum drift velocity, $v_d^{max}$, is achieved when $\omega_B=\nu$, i.e. when a significant fraction of the electrons complete whole Bloch oscillations before scattering. As $\omega_B$ increases beyond $\nu$, $v_d^{max}$ decreases with increasing $T$ according to the equation
\begin{equation}\label{eq:MaxDriftVelocityT>0Theta=0}
v_d^{max}(T)=v_0\frac{\displaystyle I_1\left(\frac{\Delta}{2k_B T}\right)}{\displaystyle I_0 \left(\frac{\Delta}{2k_BT}\right)}.
\end{equation}

The variation of $v_d^{max}$ with $T$ predicted by Eq. (\ref{eq:MaxDriftVelocityT>0Theta=0}) is shown by the solid curve in Fig. \ref{fig:vdB0}(b). For comparison, the result of our numerical simulation is shown by the symbols ``$\bullet$''. The figure shows excellent quantitative agreement between our analytical and numerical results.

\section{Effect of temperature on electron drift velocity in a tilted magnetic field}
\label{sec:T_vd}
We have also applied the numerical procedure described in the previous section to the case of a tilted magnetic field, for which the drift velocity $v_d$ cannot be determined analytically.
\begin{figure}[t]
  \centering
\includegraphics*[width=.8\linewidth]{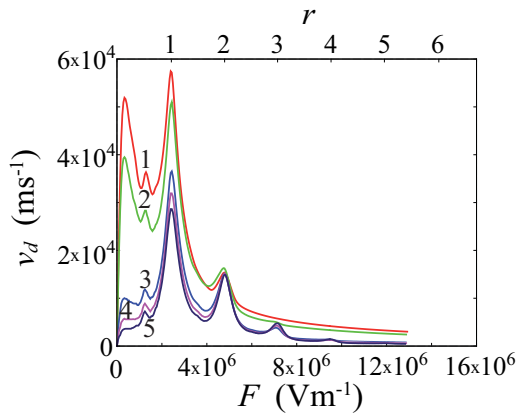}
 \caption{(Color online) Variation of drift velocity, $v_d$, with (lower scale) electric field, $F$, (upper scale) $r=\omega_B/\omega_{||}$ calculated for $B=15$ T and $\theta=40^{\circ}$ and different temperatures. Curve~1 corresponds to $T=0$~K; curve~2 to $T=50$~K; curve~3 to $T=200$~K; curve~4 to $T=300$~K; curve~5 to $T=400$~K.
\label{fig:vdB15T40}}
\end{figure}
Figure \ref{fig:vdB15T40} shows $v_d$ versus $F$ curves calculated numerically for $B=15$ T and $\theta=40^{\circ}$ at several different temperatures $T$.

In contrast to the case of $B=0$ [Fig. \ref{fig:vdB0}(a)], all of the $v_d(F)$ curves in Fig. \ref{fig:vdB15T40} exhibit multiple maxima. The first maximum, for the lowest value of $F$, also exists when $B=0$ and, as noted in the previous section, is associated with the onset of Bloch oscillations. All other maxima occur because of the enhanced acceleration of the electrons whenever the ratio of the Bloch and cyclotron frequencies $r=\omega_B/\omega_{\perp}=0.5, 1, 2$, or $3$ (upper scale in Fig. \ref{fig:vdB15T40}) \cite{From01,From04,Balanov:2008_SL-PRE}. Remarkably, as $T$ increases, the first peak weakens significantly, whereas the amplitude of the cyclotron-Bloch resonances (i.e. the peak-to-valley ratio) increases sharply even though the peak $v_d$ values decrease. Moreover, for high enough temperatures, new peaks, reflecting higher order resonances,  appear in the $v_d(F)$ dependencies.
\begin{figure}[t]
  \centering
\includegraphics*[width=.7\linewidth]{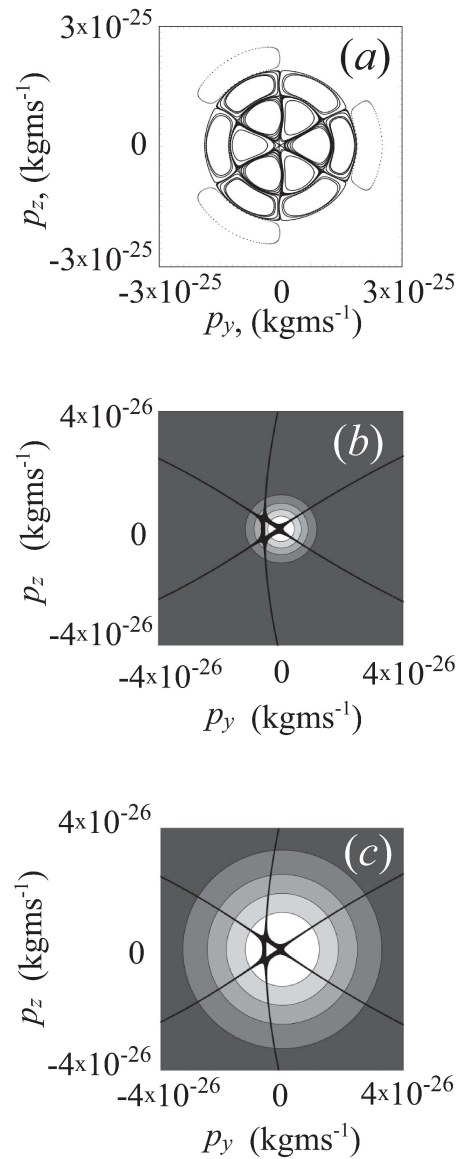}
 \caption{Stroboscopic Poincar\'e section of electron trajectories in the $(p_y,p_z)$ plane taken with strobe period $T_B=2\pi/\omega_B$ for the $r=\omega_B/\omega_{||}=3$ resonance ($B=15$ T, $\theta=40^{\circ}$). (a) ``Stochastic web'' and neighboring quasi-periodic  orbits; (b), (c)  enlarged part of ``stochastic web'' (black dots) and probability (grey-scale map in which lighter shades correspond to higher probability) that the initial momentum ($P_y, P_z$) lies in the given area of the $(p_y,p_z)$ plane calculated for $T=50$ K (b), and $T=400$ K (c).
\label{fig:webz}}
\end{figure}

To qualitatively understand the effects of temperature on drift velocity, we analysed single electron orbits described by the Eqs. (\ref{compeq1})-(\ref{compeq3}). Figure \ref{fig:webz}(a) shows a stroboscopic Poincar\'e section phase portrait for the $r=\omega_B/\omega_{||}=3$ resonance ($B=15$ T and $\theta=40^{\circ}$), taking the strobe period to be $T_B=2\pi/\omega_B$. In addition to localised quasi-periodic orbits, which appear in the stroboscopic section as dotted closed curves, the phase space also contains a resonant unbounded structure formed by chaotic layers and known in the literature as a ``stochastic web'' \cite{Sag88,Zasl91}. It has been shown previously \cite{From01,From04} that electrons having initial momenta in the stochastic web move rapidly through the SL, so producing a large resonant peak in the $v_d(F)$ curve, as shown in Fig. \ref{fig:vdB15T40} for $r=0.5,1,2,3,4$. The height of each resonant peak increases as more electrons have momenta lying within the stochastic web. In Figs. \ref{fig:webz}(b),(c) we show enlarged fragments of the stochastic web near the origin together with a grey-scale map indicating the probability of finding an electron with given initial momentum components $(P_y, P_z)$ at $T=50$ K [Fig. \ref{fig:webz} (b)] and $T=400$ K [Fig. \ref{fig:webz}(c)]. Comparison of these two figures reveals that, with increasing temperature, the probability of the electron's initial momentum lying within the stochastic web grows. At the same time, the number of electrons with higher initial momenta, and thus  higher velocity, also increases. Together, these two effects increase $v_d$ and thus enhance the Bloch - cyclotron resonant peaks in Fig.~\ref{fig:vdB15T40}.

Interestingly, although Eq. (\ref{eq:Bass}) is unable correctly to predict the exact electron drift velocity for an arbitrary magnetic field configuration, it still captures certain qualitative trends in the variation of $v_d$ with the field parameters and $T$. First, we note that according to Eq. (\ref{eq:Bass}), the expression for $v_d$ consists of a series of resonant terms proportional to $\nu(\omega_B\pm n\omega_{||})/[\nu^2+(\omega_B\pm n\omega_{||})^2]$. The term with $n=0$ reflects the Esaki-Tsu $v_d(F)$ curve, whereas other terms with $n\ne0$ correspond to the Bloch - cyclotron resonances, for which $\omega_B/\omega_{||}=n$ is an integer. Moreover, since, for very large $x$, $I_n(x)$ has the asymptotic form $I_n(x)\varpropto \exp(x)/\sqrt{x}$, as $T$ increases in Eq. (\ref{eq:Bass}) the higher-order resonant terms strengthen. Under certain conditions this leads to the appearance of additional peaks in  $v_d(F)$, which we see in our numerical simulations shown in Fig. \ref{fig:vdB15T40}.

\section{Effect of temperature on electric current through the semiconductor superlattice.}
\label{sec:T_n}

In order to study the effects of temperature on the collective dynamics of electrons in the SL, we self-consistently solve discrete versions of the current continuity and Poisson equations, splitting the miniband transport region into $N=480$ layers, each of width $\Delta x=L/N=0.24$~nm, which is small enough to approximate a continuum \cite{Greenaway:2009_SL_B,Greenaway:2010}.

The evolution of the electron density, $n_m$, in the $m^{th}$ layer, whose right-hand edge is at $x=m\Delta x$, is given by the discretised current continuity equation
\begin{equation}
 e \Delta x \frac{d n_m}{d t} = J_{m - 1} - J_m, \ \ \ m = 1 \ldots N,  \label{eq:continuity}
\end{equation}
where $J_{m-1}$ [$J_{m}$] are the areal current densities at the left- [right-] hand edges of the $m^{th}$ layer. Within the drift-diffusion approximation, the current density is
\begin{equation}
\label{eq:J_eq}
J_m=e n_m v_d\left(\overline{F}_m\right)+D\left(\overline{F}_m\right)\frac{n_{m+1}-n_m}{\Delta x}
\end{equation}
where $\overline{F}_m$ is the mean electric field in the $m^{th}$ layer\cite{Greenaway:2009_SL_B},  
\begin{equation}
\label{eq:Diff_eq}
D\left(\overline{F}_m\right)=\frac{v_d\left(\overline{F}_m\right) d}{1-\exp\left(-\frac{e \overline{F}_m d}{kT}\right)}\exp\left(-\frac{e \overline{F}_m d}{kT}\right)
\end{equation}
is the diffusion coefficient \cite{WAC02}, and the drift velocity $v_d\left(\overline{F}_m\right)$ is obtained by using the approach described in Section \ref{sec:num_proc}. Since $J_m$ depends on the local drift velocity, $v_d\left(\overline{F}_m\right)$, the collective electron dynamics depend directly on the single-electron orbits.

The electric field $F_m$ [$F_{m+1}$] at the left- [right-] hand edges of the $m^{th}$ layer can be described by the discretised Poisson equation
\begin{equation}
F_{m + 1} = \frac{e \Delta x}{\varepsilon_0 \varepsilon_r} \left( n_m - n_D \right) + F_m, \ \ \ m = 1 \ldots N,  \label{eq:poisson} \\
\end{equation}
where $\varepsilon_0$ and $\varepsilon_r=12.5$ are, respectively, the absolute and relative permittivities and $n_D=3\times10^{22}$~m$^{-3}$ is the n-type doping density in the SL layers\cite{From04}. In the emitter and collector Ohmic contacts, $F=F_0$.

\begin{figure}[tb]
  \centering
\includegraphics*[width=.7\linewidth]{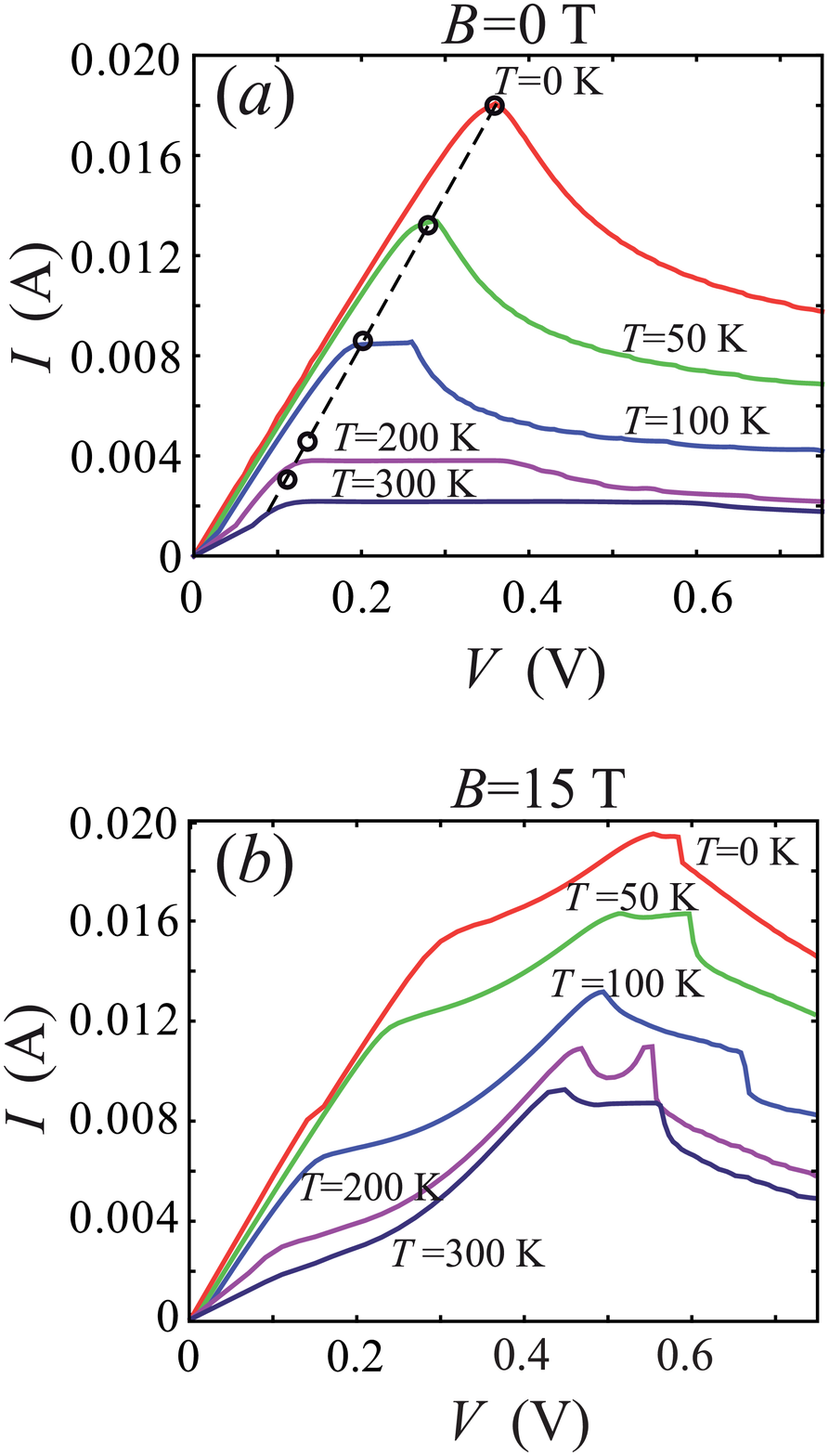}
 \caption{(Color online) $I(V)$ characteristics calculated for
${B=0}$~T (a) and ${B=15}$~T, ${\theta=40^\circ}$ (b) at (from top to bottom) $T=0$ K, 50 K, 100 K, 200 K and 300 K. In (a), data points on dashed line show values of $(V_{th}, I_{th})$ calculated from Eqs. (\ref{eq:CriticalCurrent}) and (\ref{eq:CriticalVoltage}) and discussed in the text. 
\label{fig:IvsV}}
\end{figure}

We use Ohmic boundary conditions to determine the current, $J_0=\sigma F_0$, in the heavily doped emitter contact whose electrical conductivity $\sigma=3788$~Sm$^{-1}$.\cite{From04} The voltage, $V$, applied to the device is a global constraint given by
\begin{equation}
V = U + \frac{\Delta x}{2} \sum_{m = 1}^N (F_m + F_{m + 1}),  \label{eq:volt} \\
\end{equation}
where the voltage, $U$, dropped across the contacts includes the effect of charge accumulation and depletion in the emitter and collector regions and a contact resistance $R=17~\Omega$\cite{com1}. We calculate the current as
\begin{equation}
I(t) = \frac{A}{N+1} \sum_{m = 0}^N J_m, \label{eq:tot_cur}
\end{equation}
where $A=5\times10^{-10}$~m$^2$ is the cross-sectional area of the SL.\cite{WAC02,From04,Greenaway:2009_SL_B}

The model described by Eqs. (\ref{eq:continuity}) - (\ref{eq:tot_cur}) exhibits both constant and oscillating electric current, depending on the voltage, $V$, applied to the device. Typical current-voltage [$I(V)$] characteristics are shown in Fig. \ref{fig:IvsV} both for $B=0$ and when a tilted magnetic field ($B=15$~T, $\theta=40^\circ$) is applied. Note that for $V$ values at which current oscillations occur, the DC current was calculated by averaging $I(t)$ over time. 

When $B=0$ [Fig. \ref{fig:IvsV} (a)], the $I(V)$ curves reveal the usual Esaki-Tsu-like behavior, characterised by a single maximum, which is associated with the onset of single-electron Bloch oscillations throughout much of the SL transport region. Figure \ref{fig:IvsV} (a) shows that as $T$ increases, the peak current decreases, which agrees well with a number of experimental observations \cite{Brozak90,Goutiers95}. This decrease of the peak current reflects that of the maximal electron drift velocity, predicted by Eq. (\ref{eq:MaxDriftVelocityT>0Theta=0}), as $T$ increases. In particular, the factor $I_1(\Delta/2k_BT)/I_0(\Delta/2k_BT)$, characterising the effect of the temperature on the drift velocity in Eq. (\ref{eq:MaxDriftVelocityT>0Theta=0}), also accurately describes the drop of the peak current both in our simulations and in earlier experiments discussed in Refs. \cite{Brozak90,Goutiers95}

The electron dynamics changes significantly when a tilted magnetic field is applied to the SL. Figure \ref{fig:IvsV} (b) shows typical $I(V)$ curves calculated, in the presence of a tilted magnetic field, for a range of temperatures. All of the $I(V)$ curves reveal clear Bloch-cyclotron resonances, which manifest themselves through the appearance of additional features in the curves \cite{From04}. For low temperatures, the Bloch-cyclotron resonances produce sudden changes in the slope of the $I(V)$ curves [for example, the kink near $V=0.3$ V in the (upper) $I(V)$ curve calculated for $T=0$ in Figure \ref{fig:IvsV} (b)] and also shift the position of the current peak. As $T$ increases, these effects become more prominent and, eventually, give rise to additional maxima in the $I(V)$ curves, e.g. the double peaks at $T=200$ K. This evolution of the $I(V)$ characteristics originates from the variation of the $v_d(F)$ curves with changing temperature, shown in Fig. \ref{fig:vdB15T40}. 
Comparison of Figs. \ref{fig:IvsV} (a) and (b) shows that although the maximum current decreases with increasing temperature in both cases, the peak DC current is always larger in the presence of a tilted magnetic field. 
This can be explained by the strong resonant enhancement of $v_d$ produced by the Bloch-cyclotron resonances.
Moreover, as Fig. \ref{fig:vdB15T40} indicates, with increasing temperature the amplitude of these resonances increases and so the difference between the peak currents for $B=0$ and for $B\ne0$ becomes larger. 

\begin{figure}[tb]
  \centering
\includegraphics*[width=.7\linewidth]{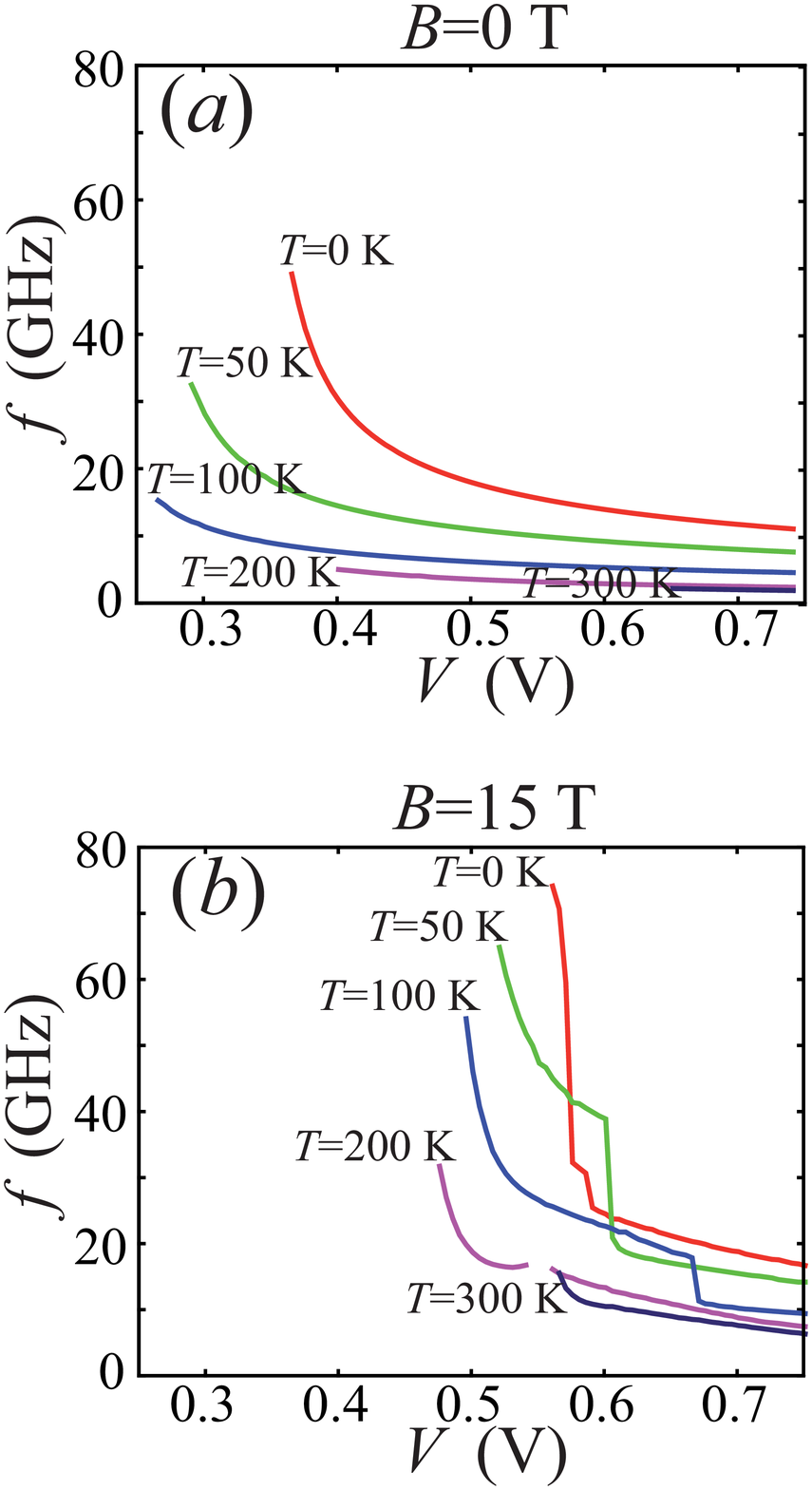}
 \caption{(Color online) Variation of the generation frequency, $f$, with voltage, $V$, applied to the SL calculated for $B=0$~T (a) and $B=15$~T,
$\theta=40^\circ$ (b) at the temperatures indicated.
\label{fig:FvsV}}
\end{figure}

When the applied voltage, $V$, exceeds some critical value, $V_{th}$, which depends on $T$, $B$, and $\theta$, the stationary state of the system loses its stability, and the electric current starts to oscillate at a frequency, $f$, which, for the given parameters, is in the microwave range. These current oscillations are associated with the formation of traveling charge domains \cite{WAC02,Greenaway:2009_SL_B}. 
\begin{figure}[tb]
  \centering
\includegraphics*[width=.7\linewidth]{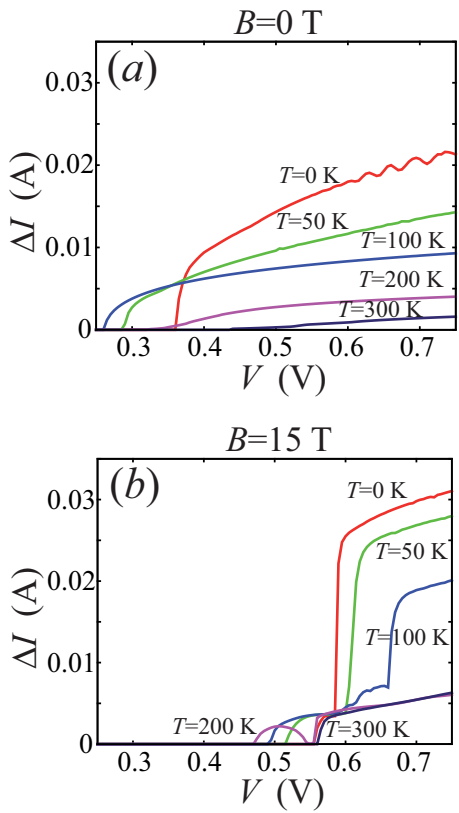}
 \caption{(Color online) Variation of the amplitude of the current oscillation, $\Delta I$, with voltage, $V$, applied to the SL calculated for $B=0$~T (a) and $B=15$~T, $\theta=40^\circ$ (b) at the temperatures indicated.
\label{fig:IAvsV}}
\end{figure}

The variation of the oscillation frequency, $f$, with $V$ is shown in Fig.~\ref{fig:FvsV} for a range of $T$. Note that for each $T$ value, the corresponding $f(V)$ curve starts at a particular threshold voltage, $V_{th}$, above which $I(t)$ oscillations occur. In general, as $V$ increases, $f$ decreases for both $B=0$ [Fig.~\ref{fig:FvsV} (a)] and $B=15$ T, $\theta=40^\circ$ [Fig.~\ref{fig:FvsV} (b)]. This can be understood in terms of the $v_d(F)$ curves shown in Figs. \ref{fig:vdB0} and \ref{fig:vdB15T40}. These figures suggest that when $F$ is sufficiently large, $v_d$ decreases with increasing $F$ for both $B=0$ and $B=15$ T. Moreover, further increasing $F$ in this regime slows the electrons even more. The applied voltage produces an inhomogeneous electric field distribution throughout the SL layers, which is determined by Eqs. (\ref{eq:continuity})-(\ref{eq:volt}). As $V$ increases, the region of high electric field inside the SL, where $v_d$ is small, also increases and so the speed of the charge domains is low. The larger the region of high electric field, the smaller the average velocity of the charge domains and, hence, the lower the frequency of current oscillations. When $B=15$ T and $\theta=40^\circ$, the multiple peaks in the $v_d(F)$ curve tend to keep the average $v_d$ value high, which means that the charge domains move faster than for $B=0$. In addition, the onset of Bloch-cyclotron resonances increases the \emph{maximal} $v_d$ value that the electrons attain for given $T$ [see Fig. \ref{fig:vdB15T40}]. All of these factors improve the propagation speed of charge domains in the presence of a tilted magnetic field. Thus, for given $V$, the frequency of charge domain oscillations (if they occur at that $V$) is higher when a tilted magnetic field is applied. 

The evolution of the amplitude of the current oscillations, $\Delta I$, with $V$ is shown in Fig.~\ref{fig:IAvsV}. Comparison of Figs. \ref{fig:IAvsV} (a) and \ref{fig:IAvsV} (b), reveals that applying a tilted magnetic field at any temperature allows one to generate current oscillations of higher amplitude and power. The mechanisms that produce this power enhancement were discussed in detail in Ref.\cite{Greenaway:2009_SL_B}. In particular, it was shown that multiple peaks in the $v_d(F)$ curves induce multiple propagating charge domains, which strengthen the current oscillations. Since the multi-peak character of the $v_d(F)$ curves persists over a wide range of $T$ (see Fig. \ref{fig:vdB15T40}), the power enhancement induced by a tilted magnetic field occurs even at room temperature. 

Remarkably, the modification of the $I(V)$ curves by a tilted magnetic field can lead to the appearance of new regions of negative differential conductance, for example associated with the two peaks in the $I(V)$ curve at $T=200$ K (purple curve in Fig. \ref{fig:IvsV}). In this case, there are several distinct regions of $V$ within which the SL exhibits current oscillations [see Fig. ~\ref{fig:IAvsV} (b)].   

Figures~\ref{fig:FvsV} and \ref{fig:IAvsV} indicate that for both $B=0$ and for $B=15$ T, the threshold voltage for current oscillations, $V_{th}$, depends non-trivially on $T$. For the low temperature range $T \lesssim 100$ K, increasing $T$ decreases $V_{th}$. But as $T$ increases further, $V_{th}$ starts to increase. This is due to the effect of temperature on the current density, $J$, which is described by Eqs. (\ref{eq:J_eq}) and (\ref{eq:Diff_eq}). As noted above, the generation of oscillating current in the SL is associated with the formation of traveling change domains, which occurs when the voltage, $U_{th}^{SL}$, dropped across the active layers of the SL becomes sufficiently large \cite{WAC02,Greenaway:2009_SL_B}. Eq. (\ref{eq:volt}) shows that this voltage is the difference between the total voltage, $V$, applied to the SL and the voltage, $U$, dropped across the contacts.
The voltage, $U$, in turn, may be expressed as $U=IR+U_c$, where $U_c$ describes the effect of charge accumulation and depletion in the emitter and collector regions. Thus, any changes in the maximal value of $J$ and, therefore, of $I$, will affect $U$, and thus shift the threshold voltage $V_{th}$. For low temperatures, when electron diffusion can be neglected, increasing $T$ decreases the maximal electron drift velocity and, thus, the maximal value of $I$. This reduces the voltage, $U$, dropped across the contacts, since this is proportional to the current through those contacts. When $U$ decreases but $V$ remains constant, the voltage, $U_{th}^{SL}$, dropped across the active layers of the SL increases. Thus, within some temperature range, increasing $T$ decreases $V_{th}$. 

The value of $V_{th}$ and the corresponding critical DC current $I_{th}=I(V_{th})$ can be estimated analytically at low $T$ and when $B=0$. If one neglects electron diffusion and uses the drift velocity given by Eq.~(\ref{eq:MaxDriftVelocityT>0Theta=0}), the critical current, $I_{th}$, at temperature $T$ can be estimated as
 \begin{equation}\label{eq:CriticalCurrent}
I_{th}(T)=I_{th}^0\frac{I_1(\Delta/2k_BT)}{I_0(\Delta/2k_BT)},
\end{equation}
where $I_{th}^0$ is the critical current for $T=0$~K.
According to Eq.~(\ref{eq:volt}), the critical voltage, $V_{th}(T)$, required for current oscillations is
\begin{equation}
V_{th}(T) = U_c+I_{th}(T)R +U_{th}^{SL},
\end{equation}
where 
\begin{equation}
U_{th}^{SL}= \frac{\Delta x}{2} \sum_{m = 1}^N (F_m^{th} + F_{m +1}^{th})
\end{equation}
is the voltage dropped across the SL layers. Assuming that $U_c$ and $U_{th}^{SL}$ are almost independent of $T$, one obtains
\begin{equation}\label{eq:CriticalVoltage}
V_{th}(T)=V_{th}^0-I_{th}^0R\left(1-\frac{I_1(\Delta/2k_BT)}{I_0(\Delta/2k_BT)
} \right),
\end{equation}
where $V_{th}^0$ is the critical voltage at $T=0$. 
From Eqs. (\ref{eq:CriticalCurrent}) and (\ref{eq:CriticalVoltage}) it follows that 
\begin{equation}\label{eq:CriticalLine}
V_{th}(T)=V_{th}^0+R(I_{th}(T)-I_{th}^0), \quad (I_{th}(T)<I_{th}^0).
\end{equation}

The $I_{th}(V_{th})$ curve defined by Eq.~(\ref{eq:CriticalLine}) is shown by the dashed line in Fig.~\ref{fig:IvsV}~(a). The open circles mark the coordinates $(V_{th}(T),I_{th}(T))$ for the five different $T$ values corresponding to the five $I(V)$ plots shown in the figure. The values of $V_{th}^0$ and $I_{th}^0$ were taken from the $I(V)$ curve for $T=0$. From Figs. ~\ref{fig:IvsV}~(a) and \ref{fig:FvsV}(a) it is easy to see that for the low $T$ values considered ($\leq 300$ K), the analytical values of $V_{th}$ and $I_{th}$ obtained from Eq. (\ref{eq:CriticalLine}) are in excellent quantitative agreement with those inferred from the numerical $I(V)$ curves.

For large $T$, the diffusion contribution to the current becomes important and Eq. (\ref{eq:CriticalLine}) fails. As $T$ increases, the current through the SL and the voltage dropped across the contacts also increase, thereby raising $V_{th}$.

The same mechanism operates when $B\ne0$. In this case, $V_{th}$ is larger than for $B=0$ (see Fig. \ref{fig:FvsV}). This is because in the presence of a tilted magnetic field, the maximal drift velocity is larger due to the onset of Bloch-cyclotron resonances (compare Figs. 3 and 4). 

\section{Conclusions}

In conclusion, we have shown that a tilted magnetic field strongly affects, and can significantly enhance, the transport characteristics of SLs -- even at room temperature. The temperature dependence of the electron drift velocity in the presence of a tilted magnetic field is quite different from the $B=0$ case. In particular, increasing $T$ quickly suppresses the $B=0$ Esaki-Tsu peak in the $v_d(F)$ curve but has a much smaller effect when a tilted magnetic field is applied. In this geometry, increasing $T$ sometimes even enhances the drift velocity peaks, relative to the background, caused by the Bloch-cyclotron resonances.  As a result, when $B$ and $\theta$ are both non-zero, the electrons move faster, which increases both the DC- and AC-components of the current through the SL and also improves its high-frequency performance. The effects of temperature on the $v_d(F)$ curves also transform the SL's $I(V)$ curves by inducing new instabilities and transport regimes, which we hope will stimulate future experimental and theoretical studies. Since the effects that we have demonstrated are generic features of semiclassical energy band transport, we expect similar phenomena to occur in, for example, cold atoms \cite{SCO02,PON06} and photonic crystals \cite{WIL03}.

\section{Acknowledgements}
The work was supported by the Federal special purpose programme ``Scientific
and educational personnel of innovation Russia (2009 - 2013)'', the UK
Engineering and Physical Sciences Research Council and the
``Dynasty'' Foundation.

\section{Appendix}
\label{sec:App}

We solved Eqs. (\ref{compeq1})-(\ref{compeq3}) 
assuming $\omega_{||}\gg\omega_{\perp}$. Under this condition
Eqs. (\ref{compeq1})-(\ref{compeq3}) can be linearised
as
\begin{eqnarray}
  \dot{p}_x (t) & = & eF - \omega_{\perp} p_y \label{a0eq1}\\
  \dot{p}_y (t) & = & - \text{$\text{$\omega_{\|}$}$} p_z (t)
  \label{a0eq2}\\
  \dot{p}_z (t) & = & \text{$\text{$\omega_{\|}$}$} p_y (t),
  \label{a0eq3}
\end{eqnarray}
which for initial conditions $p_x (0) = p_{\|}$, $p_y (0) = P_y$, $p_z(0) = P_z$
yields
\begin{eqnarray}
\label{eq:mom_lin}
  &  & p_x (t) = p_{\|} + e F t - \frac{\omega_{\perp}}{\text{$\omega_{\|}$}}
  p_{\perp} \sin ( \text{$\text{$\omega_{\|}$} t$} + \varphi_0) +
  \frac{\omega_{\perp}}{\text{$\omega_{\|}$}} p_{\perp} \sin \varphi_0,
  \nonumber\\
  &  & p_y (t) = p_{\perp} \cos ( \text{$\text{$\omega_{\|}$} t$} +
  \varphi_0), \\
  &  & p_z (t) = p_{\perp} \sin ( \text{$\text{$\omega_{\|}$} t$} +
  \varphi_0), \nonumber \
\end{eqnarray}
where $p_{\perp} = \sqrt{P^2_y + P^2_z}$, and $\varphi_0 = atan (P_z /
P_y)$.

Eq. (\ref{eq:mom_lin}) for the crystal momentum can be used to calculate the electron velocity along the SL ($x$) axis as a function of time, $t$:

\begin{widetext}
\begin{eqnarray}
  v_x (t) = v_0 \sin \left( \frac{p_x (t) d}{\hbar} \right) = v_0 \sin \left[
  \frac{p_{\|} d}{\hbar} + \omega_B t -
  \frac{\omega_{\perp}}{\text{$\omega_{\|}$}} \frac{d}{\hbar} p_{\perp} \sin (
  \text{$\text{$\omega_{\|}$} t$} + \varphi_0) +
  \frac{\omega_{\perp}}{\text{$\omega_{\|}$}} \frac{d}{\hbar} p_{\perp} \sin
  \varphi_0 \right] = &  &  \nonumber\\
  \frac{v_0}{2 i} \left[ \exp \left( i \frac{p_{\|} d}{\hbar} + i \omega_B t -
  i \frac{\omega_{\perp}}{\text{$\omega_{\|}$}} \frac{d}{\hbar} p_{\perp} \sin
  ( \text{$\text{$\omega_{\|}$} t$} + \varphi_0) + i
  \frac{\omega_{\perp}}{\text{$\omega_{\|}$}} \frac{d}{\hbar} p_{\perp} \sin
  \varphi_0 \right) - \right. &  & \\
  \left. \exp \left( - i \frac{p_{\|} d}{\hbar} - i \omega_B t + i
  \frac{\omega_{\perp}}{\text{$\omega_{\|}$}} \frac{d}{\hbar} p_{\perp} \sin (
  \text{$\text{$\omega_{\|}$} t$} + \varphi_0) - i
  \frac{\omega_{\perp}}{\text{$\omega_{\|}$}} \frac{d}{\hbar} p_{\perp} \sin
  \varphi_0 \right) \right], &  &  \nonumber
\end{eqnarray}
\end{widetext}
where $\omega_B = e F d / \hbar$ is frequency of the Bloch oscillations and $v_0 = \Delta d / (2 \hbar)$ is the maximal velocity in the given miniband.

Using the Jacobi-Anger expansion for the Bessel functions we obtain
\begin{widetext}
\begin{eqnarray}
\label{eq:app_vx}
  &  & v_x (t) = \frac{v_0}{2 i} \left[ e^{i \frac{p_{\|} d}{\hbar}}
  \sum^{\infty}_{n = - \infty} \sum^{\infty}_{k = - \infty} J_k \left(
  \frac{\omega_{\perp}}{\text{$\omega_{\|}$}} \frac{d}{\hbar} p_{\perp}
  \right) e^{i (\omega_B - k \omega_{\|}) t} J_n \left(
  \frac{\omega_{\perp}}{\text{$\omega_{\|}$}} \frac{d}{\hbar} p_{\perp}
  \right) e^{i \varphi_0 (n - k)} \right.\\ \nonumber
  &  & \\
  &  & \left. - e^{- i \frac{p_{\|} d}{\hbar}} \sum^{\infty}_{n = - \infty}
  \sum^{\infty}_{k = - \infty} J_k \left(
  \frac{\omega_{\perp}}{\text{$\omega_{\|}$}} \frac{d}{\hbar} p_{\perp}
  \right) e^{- i (\omega_B - k \omega_{\|}) t} J_n \left(
  \frac{\omega_{\perp}}{\text{$\omega_{\|}$}} \frac{d}{\hbar} p_{\perp}
  \right) e^{- i \varphi_0 (n - k)} \right]. \nonumber
\end{eqnarray}
\end{widetext}
The drift velocity, $u_d(p_{\|},p_{\perp})$, of electrons with initial momentum components $p_{\|}$ and $p_{\perp}$ can be calculated using the Esaki-Tsu formula (\ref{eq:ud}). Substituting (\ref{eq:app_vx}) in (\ref{eq:ud}) we get
\begin{widetext}
\begin{eqnarray}
  u_d (p_{\|}, p_{\perp,}, \varphi_0)
  = \frac{v_0}{2 i}\left[A (p_{\perp}, \varphi_0) e^{i \frac{p_{\|} d}{\hbar}} - B (p_{\perp},
  \varphi_0) e^{- i \frac{p_{\|} d}{\hbar}}\right], \label{eq:app_ud}
\end{eqnarray}
\end{widetext}
with
\begin{widetext}
\begin{eqnarray*}
  A (p_{\perp}, \varphi_0) = \sum^{\infty}_{n = - \infty} \sum^{\infty}_{k = -
  \infty} J_k \left( \frac{\omega_{\perp}}{\text{$\omega_{\|}$}}
  \frac{d}{\hbar} p_{\perp} \right) J_n \left(
  \frac{\omega_{\perp}}{\text{$\omega_{\|}$}} \frac{d}{\hbar} p_{\perp}
  \right) e^{i \varphi_0 (n - k)} \frac{\nu}{\nu - i (\omega_B - k
  \omega_{\|})} & , & \\
  B (p_{\perp}, \varphi_0) = \sum^{\infty}_{n = - \infty} \sum^{\infty}_{k = -
  \infty} J_k \left( \frac{\omega_{\perp}}{\text{$\omega_{\|}$}}
  \frac{d}{\hbar} p_{\perp} \right) J_n \left(
  \frac{\omega_{\perp}}{\text{$\omega_{\|}$}} \frac{d}{\hbar} p_{\perp}
  \right) e^{- i \varphi_0 (n - k)} \frac{\nu}{\nu + i (\omega_B - k
  \omega_{\|})} & . &
\end{eqnarray*}
\end{widetext}

To find the drift velocity, $v_d$, at finite temperature $T$ we average $u_d (p_{\|}, p_{\perp,},
\varphi_0)$ according to the thermal distribution of electron momenta (\ref{eq:Boltzmann}). In the current notations the definition of $v_d$ becomes
\begin{widetext}
\begin{eqnarray}
  v_d = \int^{\pi \hbar / d}_{- \pi \hbar / d} \int^{\infty}_{- \infty}
  \int^{\pi}_{- \pi} &  & \text{$u_d (p_{\|}, p_{\perp,}, \varphi_0)$} f
  (p_{\|}, p_{\perp})  p_{\perp} d p_{\|} d p_{\perp} d \varphi_0,
\end{eqnarray}
\end{widetext}
with
\[ f (p_{\|}, p_{\perp}) = \frac{1}{Z} \exp \left[ - \frac{\Delta}{2 k_B T}
   \left( 1 - \cos \frac{p_{\|} d}{\hbar} \right) - \frac{ p^2_{\perp} }{2
   m^{\ast} k_B T} \right] , \]
and $Z$  is given by Eq.(\ref{eq:Znorm})

To determine $v_d$ from the above integral expression, we first evaluate the integral over $p_{\|}$, which can be written as
\begin{widetext}
\begin{eqnarray*}
  u_d (p_{\perp}, \varphi_0) = \frac{1}{Z} \frac{v_0}{2 i} A (p_{\perp},
  \varphi_0) e^{- \frac{\Delta}{2 k_B T}} \int^{\pi \hbar / d}_{- \pi \hbar / d}
  \exp \left( i \frac{p_{\|} d}{\hbar} \right) \exp \left[ \frac{\Delta \cos
  \frac{p_{\|} d}{\hbar}}{2 k_B T} \right] d p_{\|} -&  & \\
  - \frac{1}{Z} \frac{v_0}{2 i} B (p_{\perp}, \varphi_0) e^{-
  \frac{\Delta}{2 k_B T}} \int^{\pi \hbar / d}_{- \pi \hbar / d} \exp \left( - i
  \frac{p_{\|} d}{\hbar} \right) \exp \left[ \frac{\Delta \cos \frac{p_{\|}
  d}{\hbar}}{2 k_B T} \right] d p_{\|} &  &
\end{eqnarray*}
or
\begin{eqnarray*}
  u_d (p_{\perp}, \varphi_0) = \frac{1}{Z} \frac{v_0}{2 i} A (p_{\perp},
  \varphi_0) e^{- \frac{\Delta}{2 k_B T}} \int^{\pi \hbar / d}_{- \pi \hbar / d}
  \sum^{\infty}_{n = - \infty} I_n \left( \frac{\Delta}{2 k_B T} \right) e^{i
  \frac{d}{\hbar} (n + 1) p_{\|}} d p_{\|} - &  & \\
  - \frac{1}{Z} \frac{v_0}{2 i} B (p_{\perp}, \varphi_0) e^{-
  \frac{\Delta}{2 k_B T}} \int^{\pi \hbar / d}_{- \pi \hbar / d} \sum_{n = -
  \infty}^{\infty} I_n \left( \frac{\Delta}{2 k_B T} \right) e^{i
  \frac{d}{\hbar} (n - 1) p_{\|}} d p_{\|} = &  & \\
  = \frac{2 \pi}{Z} \frac{v_0}{2 i} \frac{\hbar}{d} e^{- \frac{\Delta}{2 k_B T}}
  I_1 \left( \frac{\Delta}{2 k_B T} \right) \left[ A (p_{\perp}, \varphi_0) -
  B (p_{\perp}, \varphi_0) \right] &  &
\end{eqnarray*}
\end{widetext}

At the next step we perform integration over $p_{\perp}$ and $\varphi_0$
\begin{eqnarray*}
  &  & v_d = \int^{\infty}_0 \int^{\pi}_{- \pi} u_d (p_{\perp}, \varphi_0) \exp
  \left[ - \frac{p^2_{\perp}}{2 m^{\ast} k_B T} \right] p_{\perp} d p_{\perp}
  d \varphi_0
\end{eqnarray*}

\begin{widetext}
\begin{eqnarray}
  v_d = &  & \frac{2 \pi}{Z} \frac{v_0}{2 i} \frac{\hbar}{d} e^{-
  \frac{\Delta}{2 k_B T}} I_1 \left( \frac{\Delta}{2 k_B T} \right)
  \int^{\infty}_0 \int^{\pi}_{- \pi} \sum^{\infty}_{n = - \infty}
  \sum^{\infty}_{k = - \infty} J_k \left(
  \frac{\omega_{\perp}}{\text{$\omega_{\|}$}} \frac{d}{\hbar} p_{\perp}
  \right) J_n \left( \frac{\omega_{\perp}}{\text{$\omega_{\|}$}} \nonumber
  \frac{d}{\hbar} p_{\perp} \right) e^{- i \varphi_0 (n - k)}\\ \nonumber
  &  & \times \left[ \frac{\nu}{\nu - i (\omega_B - k \omega_{\|})} -
  \frac{\nu}{\nu + i (\omega_B - k \omega_{\|})} \right] \exp \left[ -
  \frac{p^2_{\perp}}{2 m^{\ast} k_B T} \right] p_{\perp} d p_{\perp} d
  \varphi_0=\\ \nonumber
  = &  & \frac{(2 \pi)^2 v_0}{Z} \frac{\hbar}{d} e^{- \frac{\Delta}{2 k_B
  T}} I_1 \left( \frac{\Delta}{2 k_B T} \right) \int^{\infty}_0
  \sum^{\infty}_{n = - \infty} J^2_n \left(
  \frac{\omega_{\perp}}{\text{$\omega_{\|}$}} \frac{d}{\hbar} p_{\perp}
  \right) \frac{\nu (\omega_B - n \omega_{\|})}{\nu^2 + (\omega_B - n
  \omega_{\|})^2} \times \exp \left[ - \frac{p^2_{\perp}}{2 m^{\ast} k_B T} \right]
  p_{\perp} d p_{\perp},\\  \label{vdprel}
\end{eqnarray}
\end{widetext}
Taking into account that
$J_{- n} (x) = (- 1)^n
J_n (x)$, we can rewrite (\ref{vdprel})
\begin{widetext}
\begin{eqnarray}
\label{eq:app_vd2}
  v_d  = \frac{1}{2} \frac{(2 \pi)^2 v_0}{Z} \frac{\hbar}{d} e^{-
  \frac{\Delta}{2 k_B T}} I_1 \left( \frac{\Delta}{2 k_B T} \right)
  \sum^{\infty}_{n = 0} \left[ \int^{\infty}_0 J^2_n \left(
  \frac{\omega_{\perp}}{\text{$\omega_{\|}$}} \frac{d}{\hbar} p_{\perp}
  \right) \frac{\nu (\omega_B - n \omega_{\|})}{\nu^2 + (\omega_B - n
  \omega_{\|})^2} \right.\times \exp \left[ - \frac{p^2_{\perp}}{2 m^{\ast} k_B T} \right]
  p_{\perp} d p_{\perp} +\\ \nonumber
   \left. + \int^{\infty}_0 J^2_n \left(
  \frac{\omega_{\perp}}{\text{$\omega_{\|}$}} \frac{d}{\hbar} p_{\perp}
  \right) \frac{\nu (\omega_B + n \omega_{\|})}{\nu^2 + (\omega_B - n
  \omega_{\|})^2} \times \exp \left[ - \frac{p^2_{\perp}}{2 m^{\ast} k_B T} \right]
  p_{\perp} d p_{\perp} \right].
\end{eqnarray}
\end{widetext}
The integral
\begin{widetext}
\begin{eqnarray}
  \int^{\infty}_0 J^2_n \left( \frac{\omega_{\perp}}{\omega_{\|}}
  \frac{d}{\hbar} p_{\perp} \right) \exp \left[ - \frac{p^2_{\perp}}{2
  m^{\ast} k_B T} \right] p_{\perp} d p_{\perp} =
  m^{\ast} k_B T \exp \left[ - m^{\ast} k_B T \left(
  \frac{\omega_{\perp}}{\text{$\omega_{\|}$}} \frac{d}{\hbar} \right)^2
  \right] I_n \left[ m^{\ast} k_B T \left(
  \frac{\omega_{\perp}}{\text{$\omega_{\|}$}} \frac{d}{\hbar} \right)^2
  \right] &  &
\end{eqnarray}
\end{widetext}
is, for $n > - 1$, a special from of the second Weber exponential integral \cite{Watson62}.
After integration and re-indexing the sum in (\ref{eq:app_vd2}), and substitution of the expression for $Z$ from (\ref{eq:Znorm}) we obtain the final formula for the drift velocity
\begin{widetext}
\[ v_d = v_0 \frac{I_1 (\Delta / 2 k_B T)}{I_0 (\Delta / 2 k_B T)} \exp \left[
   - m^{\ast} k_B T \left( \frac{\omega_{\perp}}{\text{$\omega_{\|}$}}
   \frac{d}{\hbar} \right)^2 \right] \sum^{\infty}_{n = - \infty} I_n \left[
   m^{\ast} k_B T \left( \frac{\omega_{\perp}}{\text{$\omega_{\|}$}}
   \frac{d}{\hbar} \right)^2 \right] \frac{\nu (\omega_B - n
   \omega_{\|})}{\nu^2 + (\omega_B - n \omega_{\|})^2}. \]
\end{widetext}


\begin{thebibliography}{9}
\bibitem{footnote1} The electron scattering rate $\nu$ can describe both elastic and inelastic processes, see
for example Ref.\cite{WAC02}.
\bibitem{com1}In order to simplify the analysis, we neglect the temperature dependence of the contact resistance and electron scattering rate, which is an acceptable assumption for the SLs used in several experiments \cite{Brozak90,Goutiers95}.
\end{thebibliography}
\end{document}